\documentclass{aa}
\usepackage{graphicx}

\def\degr{\mathrm{^\circ}}

\def\aap{A\&A}

\def\apj{ApJ}

\def\apjs{ApJS}

\begin{document}

\title{The spinning-top Be star Achernar from VLTI-VINCI}

\author{ A.~Domiciano~de~Souza \inst{1}
\and P.~Kervella \inst{2}
\and S.~Jankov \inst{3}
\and L.~Abe  \inst{1}
\and F.~Vakili \inst{1}
\and E. di Folco \inst{4}
\and F.~Paresce \inst{4}
}

\institute{Laboratoire Univ. d'Astroph. de Nice (LUAN), CNRS UMR
6525, Parc Valrose, 06108 Nice Cedex 02, France
\and European Southern Observatory (ESO), Alonso de Cordova 3107,
Casilla 19001, Vitacura, Santiago 19, Chile
\and
Observatoire de la C\^{o}te d'Azur, D\'{e}partement FRESNEL, CNRS UMR
6528, Boulevard de l'Observatoire, B.P 4229, 06304 Nice, France
\and
European Southern Observatory (ESO), Karl-Schwarzschild str. 2,
D-85748 Garching, Germany }

\offprints{A.~Domiciano~de~Souza,\email{Armando.Domiciano@obs-azur.fr}}

\date{Received $<date>$ / Accepted $<date>$}

\abstract{We report here the first observations of a rapidly
rotating Be star, $\alpha$ Eridani, using Earth-rotation synthesis
on the Very Large Telescope (VLT) Interferometer. Our measures
correspond to a $2a/2b = 1.56\pm0.05$ apparent oblate star, $2a$
and $2b$ being the equivalent uniform disc angular diameters in
the equatorial and polar direction. Considering the presence of a
circumstellar envelope (CSE) we argue that our measurement
corresponds to a truly distorted star since $\alpha$ Eridani
exhibited negligible H$\alpha$ emission during the interferometric
observations. In this framework we conclude that the commonly
adopted Roche approximation (uniform rotation and centrally
condensed mass) should not apply to $\alpha$ Eridani. This result
opens new perspectives to basic astrophysical problems, such as
rotationally enhanced mass loss and internal angular momentum
distribution. In addition to its intimate relation with magnetism
and pulsation, rapid rotation thus provides a key to the Be
phenomenon: one of the outstanding non-resolved problems in
stellar physics.  }

\maketitle

\section{Introduction}

The southern star Achernar ($\alpha$ Eridani, HD 10144, spectral
type B3Vpe) is the brightest Be star in the sky. A Be star is
defined as a non-supergiant B type star that has presented
episodic Balmer lines in emission (Jaschek et al. 1981), whose
origin is attributed to a CSE ejected by the star itself. Physical
mechanisms like non-radial pulsations, magnetic activity, or
binarity are invoked to explain the CSE formation of Be stars in
conjunction with their fundamental property of rapid rotation.
Theoretically, rotation has several consequences on the star
structure (Cassinelli 1987). The most obvious is the geometrical
deformation that results in a larger radius at the equator than at
the poles. Another well established effect, known as gravity
darkening or the von Zeipel effect for hot stars (von Zeipel
1924), is that both surface gravity and emitted flux decrease from
the poles to the equator. Although well studied in the literature,
the effects of rotation have rarely been tested against accurate
enough observations (Reiners \& Schmitt 2003, van Belle et al.
2001), a gap bridged by our interferometric observations of
Achernar.

\section{Observations and data processing}

Dedicated observations of Achernar have been carried out during
the ESO period 70, from 11 September to 12 November 2002, with
quasi-uniform time coverage, on the VLT Interferometer (VLTI,
Glindemann et al. 2003) equipped with the VINCI beam combiner
(Kervella et al. 2003). This instrument recombines the light from
two telescopes in the astronomical K band, which is centered at
2.2 $\mu$m and covers 0.4 $\mu$m. The observable measured by VINCI
is the squared coherence factor $\mu^2$ of the star light. It is
derived from the raw interferograms after photometric calibration
using a wavelet based method (S\'{e}gransan et al. 1999). The
reduction procedure is detailed by Kervella et al. (2003) and has
successfully been applied to dwarf stars observations with the
VLTI (S\'{e}gransan et al. 2003). The instrumental value of $\mu^2$ is
then calibrated through the observation of stable stars with known
angular diameters. The calibrators chosen for Achernar are
presented in Table 1. The final product of the processing is the
squared visibility $V^2$ of the object for each baseline projected
on the sky ($B_\mathrm{proj}$). $V^2$ is directly related to the
Fourier transform of the brightness distribution of the object via
the Zernike-Van Cittert theorem. For these observations, two
interferometric baselines were used, 66~m (E0-G1; azimuth
$147\degr$, counted from North to East) and 140~m (B3-M0;
$58\degr$), equipped with 40 cm siderostats (Fig. 1 left). Their
orientations are almost perpendicular to each other giving an
excellent configuration for the detection of stellar asymmetries.
Moreover, Earth-rotation has produced an efficient baseline
synthesis effect (Fig. 1 right). A total of more than 20000
interferograms were recorded on Achernar, and approximately as
many on its calibrators, corresponding to more than 20 hours of
integration. From these data, we obtained 60 individual $V^2$
estimates, at an effective wavelength of $\lambda_\mathrm{eff} =
2.175 \pm 0.003$ $\mu$m.

\begin{table}[ht]
 \centering
 \caption[]{Relevant parameters of the calibrators of
Achernar. $\oslash_\mathrm{UD}$ is the equivalent uniform disc
angular diameter. The value of $\oslash_\mathrm{UD}$ for $\delta$
Phe and $\chi$   Phe is based on spectrophotometry (Cohen et al.
1999), while that for $\alpha$ PsA  was measured separately with
the VLTI and should appear in a forthcoming publication from one
of us (E.F.).}\label{ta:calibrators}
 \vspace{0.5cm}
\begin{tabular}{ccccc}
\hline\hline

Name  &    Spec.type  &  $\lambda_\mathrm{eff}$ & Baseline &  $\oslash_\mathrm{UD}$ \\
      &             &   ($\mu$m)              &   (m)    &    (mas) \\
\hline

$\delta$ Phe  &  K0IIIb  &  2.181  &  140  &  $2.18 \pm 0.02$ \\
$\alpha$ PsA  &  A3V     &  2.177  &  140  &  $2.20 \pm 0.07$ \\
$\chi$   Phe  &  K5III   &  2.182  &  66   &  $2.69 \pm 0.03$ \\

\hline
\end{tabular}\par \vspace{0.1cm}
\end{table}
\begin{figure}[t]
 \centering
  \includegraphics*[width=9cm,draft=false]{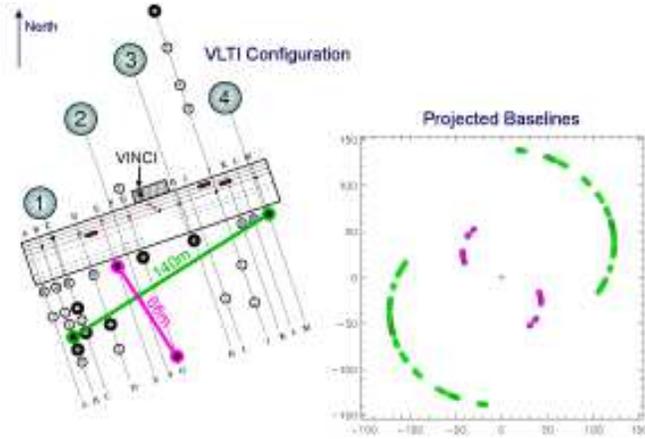} 
  \caption{VLTI ground baselines for Achernar observations and their
corresponding projections onto the sky at different observing
times. \textit{Left:} Aerial view of VLTI ground baselines for the
two pairs of 35~ccm siderostats used for Achernar observations.
Color magenta represents the 66~m (E0-G1; azimuth $147 \degr$,
counted from North to East) and green the 140~m (B3-M0; $58
\degr$). \textit{Right:} Corresponding baseline projections onto
the sky ($B_\mathrm{proj}$) as seen from the star. Note the very
efficient Earth-rotation synthesis resulting in a nearly complete
coverage in azimuth angles. }
  \label{fig:baselines}
\end{figure}

\section{Results}

Determination of the shape of Achernar from our set of $V^2$ is
not a straightforward task so that some prior assumptions need to
be made in order to construct an initial solution for our
observations. A convenient first approximation is to derive from
each $V^2$ an equivalent uniform disc (UD) angular diameter
$\oslash_\mathrm{UD}$ from the relation $V^2  = \left|
{2\mathrm{J}_1 ( z )/z} \right|^2$. Here, $z = \pi
\oslash_\mathrm{UD} \left( \alpha
\right)\mathrm{B}_{\mathrm{proj}} \left( \alpha  \right)\lambda
_{\mathrm{eff}}^{\mathrm{ - 1}}$, $\mathrm{J}_1$ is the Bessel
function of the first kind and of first order, and $\alpha$ is the
azimuth angle of $B_\mathrm{proj}$ at different observing times
due to Earth-rotation. The application of this simple procedure
reveals the extremely oblate shape of Achernar from the
distribution of $\oslash_\mathrm{UD}(\alpha)$ on an ellipse
(Fig.~2). Since $\alpha$, $B_\mathrm{proj}(\alpha)$, and
$\lambda_\mathrm{eff}$ are known much better than 1\%, the
measured errors in $V^2$ are associated only to the uncertainties
in $\oslash_\mathrm{UD}$. We performed a non-linear regression fit
using the equation of an ellipse in polar coordinates. Although
this equation can be linearized in Cartesian coordinates, such a
procedure was preferred to preserve the original, and supposedly
Gaussian, residuals distribution as well as to correctly determine
the parameters and their expected errors. We find a major axis
$2a=2.53\pm0.06$ milliarcsec (mas), a minor axis $2b=1.62\pm0.01$
mas, and a minor-axis orientation $\alpha_0=39\degr\pm1\degr$.
Note that the corresponding ratio $2a/2b=1.56\pm0.05$ determines
the equivalent star oblateness only in a first-order UD
approximation. To interpret our data in terms of physical
parameters of Achernar, a consistent scenario must be tailored
from its basic known properties, so that we can safely establish
the conditions where a coherent model can be built and discussed.

\begin{figure}[t]
 \centering
  \includegraphics*[width=\hsize,draft=false]{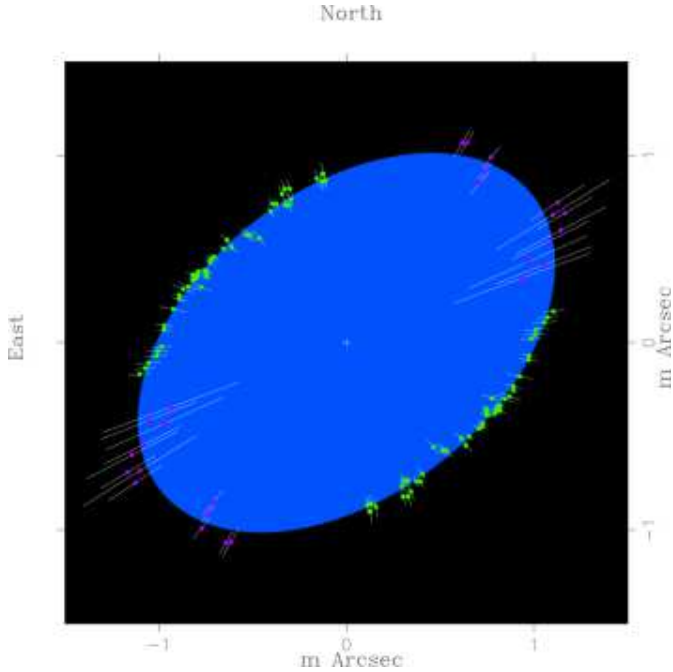}
  \caption{Fit of an ellipse over the observed squared visibilities $V^2$ translated to
equivalent uniform disc angular diameters. Each $V^2$ is plotted
together with its symmetrical value in azimuth. Magenta points are
for the 66 m baseline and green points are for the 140 m baseline.
The fitted ellipse results in major axis $2a=2.53\pm0.06$
milliarcsec, minor axis $2b~=~1.62\pm0.01$ milliarcsec, and minor
axis orientation $\alpha_0=39\degr{\pm}1\degr$ (from North to East).
The points distribution reveals an extremely oblate shape with a
ratio $2a/2b = 1.56\pm0.05$.}
  \label{fig:ellipse}
\end{figure}

\section{Discussion}

Achernar's pronounced apparent asymmetry obtained in this first
approximation, together with the fact that it is a Be star, raises
the question of whether we observe the stellar photosphere with or
without an additional contribution from a CSE.

For example, a flattened envelope in the equatorial plane would
increase the apparent oblateness of the star if it were to
introduce a significant infrared (IR) excess with respect to the
photospheric continuum. Theoretical models (Poeckert \&
Marlborough 1978) predict a rather low CSE contribution in the K
band especially for a star tilted at higher inclinations, which
should be our case as discussed below. Indeed, Yudin (2001)
reported a near IR excess (difference between observed and
standard color indices in visible and L band centered at 3.6
$\mu$m) to be E(V-L) $=0\fm2$, with the same level of uncertainty.
Moreover, this author reports a zero intrinsic polarization
($p_*$). These values are significantly smaller than mean values
for Be stars earlier than B3 ( $\overline \mathrm{E}(V-L)>0\fm5$
and $\overline {p_*}>0.6\%$), meaning that the Achernar's CSE is
weaker than in other known Be stars. Further, an intermediate
angular diameter of our elliptical fit (Fig.~2) is compatible with
the $\oslash_\mathrm{UD}=1.85\pm0.07$ mas measured by
Hanbury-Brown (1974) in the visible, in contrast to what is
expected if the envelope were to contribute to our present IR
observations. Finally, Chauville et al. (2001) report no emission
in the H$\gamma$ line. Since the emission in H$\gamma$ and in the
continuum are both formed roughly in the same layer of the CSE
(Ballereau et al. 1995), the contribution from the nearby
environment of the star should be considered below the limit of
detection.

Of course, being classified as a Be star, Achernar can enhance the
strength of its CSE due to episodic mass ejections, which are
generally witnessed by increased Balmer line emission (e.g. de
Freitas Pacheco 1982). This possibility was checked against a
H$\alpha$ spectrum (Leister \& Janot-Pacheco 2003) taken in
October 2002, during our VLTI-VINCI campaign presenting a
photospheric absorption profile. To be sure that we observed a
quiescent Achernar we synthesized a H$\alpha$ profile from our
model (Domiciano de Souza et al. 2002) compared to the observed
line. We estimated the emission to be at most 3\% across the whole
line. Such an upper limit would correspond to a CSE emitting at
most 12\% of the photospheric continuum flux, due to free-free and
free-bound emission (Poeckert \& Marlborough 1978).

Thus, we assume hereafter that the observed asymmetry of Achernar
mainly reflects its true photospheric distortion with a negligible
CSE contribution. Under this assumption, and using the Hipparcos
distance ($d=44.1\pm1.1$ pc; Perryman et al. 1997), we derive an
equatorial radius $R_\mathrm{eq}=12.0\pm0.4$ $\mathrm{R}_{\sun}$
and a maximum polar radius $R_\mathrm{pol}^\mathrm{max}=7.7\pm0.2$
$\mathrm{R}_{\sun}$, respectively from $2a$ and $2b$ obtained from
the elliptical fit on $\oslash_\mathrm{UD}(\alpha)$. From simple
geometrical considerations the actual polar radius
$R_\mathrm{pol}$ will be smaller than
$R_\mathrm{pol}^\mathrm{max}$ for polar inclinations $i <
90\degr$, while $R_\mathrm{eq}$ is independent of $i$.

\begin{table}[ht]
 \centering
 \caption[]{Fundamental parameters for two limit solution models of Achernar.
From the fixed parameters and in the Roche approximation, the
minimum polar inclination is $i_{\min} = 46\degr$ where
$V_\mathrm{eq}=V_\mathrm{crit}=311$ km/s and
$R_\mathrm{eq}=1.5R_\mathrm{pol}$. In addition to these parameters
we adopted a linear limb darkening parameter from Claret (2000)
compatible with the variable effective temperature and gravity
over the stellar surface. The equatorial effective temperature is
much lower than the polar one due to the von Zeipel
effect.}\label{ta:parameters}
 \vspace{0.5cm}
\begin{tabular}{ccc}
\hline\hline
            Fixed parameters               &        Adopted value                 &  Comments         \\
\hline\hline
        $T_\mathrm{pol}$             &   20\,000 K      &  $\sim$ B3V star \\
              Mass                  &     6.07 $\mathrm{M}_{\sun}$      &  Harmanec (1988) \\
     $V_\mathrm{eq} \sin i$         &     225 km/s                      &  Slettebak (1982)  \\
        $R_\mathrm{eq}$             &   12.0  $\mathrm{R}_{\sun}$       &  this work \\
\hline\hline
            Model dependent                 &        Values for             &  Values for       \\
            parameters                      &        Model A                &    Model B        \\
\hline\hline
       $T_\mathrm{eq}$              &       9\,500 K      &    14\,800 K\\
            $i$                     &       $50\degr$            &  $90\degr$    \\
     $V_\mathrm{crit} $         &     304 km/s                      & 285 km/s   \\
     $V_\mathrm{eq} $         &     $0.96 V_\mathrm{crit}$        &  $0.79 V_\mathrm{crit}$    \\
        $R_\mathrm{pol}$             &   8.3  $\mathrm{R}_{\sun}$       &  9.5  $\mathrm{R}_{\sun}$ \\

\hline
\end{tabular}\par \vspace{0.1cm}
\end{table}
\begin{figure}[t]
 \centering
  \includegraphics*[width=\hsize,draft=false]{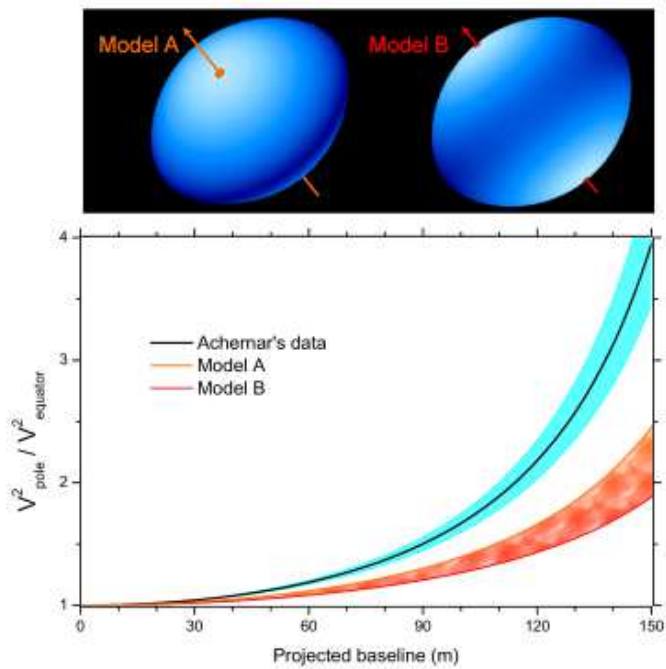}
  \caption{Comparison of ratios of squared visibility curves between
the polar and equatorial directions $V_{\mathrm{pol}}^2
/V_{\mathrm{eq}}^2$. The black solid curve corresponds to
$V_{\mathrm{pol}}^2/V_{\mathrm{eq}}^2$ for the elliptical fit on
Achernar's data, together with the corresponding uncertainties.
Simply speaking $V_{\mathrm{pol}}^2 /V_{\mathrm{eq}}^2$ somehow
reflects $R_\mathrm{eq}/R_\mathrm{pol}^\mathrm{max}$, since
interferometry is sensitive to the Fourier transform of the
stellar brightness distribution. The colored region represents our
attempt attain the black curve with our model for Achernar within
the physically reasonable solutions A (orange; upper limit) and B
(red; lower limit). This failure to reproduce the observations is
a strong and direct indication that uniform rotation does not
apply to rapidly rotating stars.}
  \label{fig:model}
\end{figure}

Based on these conclusions we applied our interferometry-oriented
code (Domiciano de Souza et al. 2002) to Achernar. This code
includes radiation transfer, the von Zeipel law ($
T_{\mathrm{eff}} \propto g_{\mathrm{eff}}^{0.25}$ ,
$T_\mathrm{eff}$ and $g_\mathrm{eff}$ being the effective
temperature and gravity, respectively), and the Roche
approximation (e.g. Roche 1837, Kopal 1987). In this approximation
and noting that the stellar rotation must be kept smaller than its
critical value, the adopted projected equatorial velocity
$V_\mathrm{eq} \sin i = 225$ km/s (Slettebak 1982) implies that $i
> 46\degr$. At this limit $T_\mathrm{eff}$ and $g_\mathrm{eff}$
both attain zero at the equator, and the surface equipotential
first derivatives become discontinuous. Therefore we chose to
explore a parameter space between the representative limit
solution models A ($i = 50\degr$) and B ($i = 90\degr$). Table 1
summarizes the corresponding sets of fundamental parameters.
Fig.~3 clearly shows that the solutions enclosed between the
models A and B cannot reproduce the observed highly oblate
ellipse. We also checked, with negative result, whether the
situation would improve significantly by varying the fundamental
parameters of Achernar in a physically reasonable range (mass $\pm
1 \mathrm{M}_{\sun}$, $T_\mathrm{pol} \pm 2000$K, $V_\mathrm{eq}
\sin i \pm 25$ km/s).

Thus, in absence of H$\alpha$ emission making a CSE contribution
unlikely to reproduce the observed oblateness, the classical
assumption of Roche approximation becomes questionable. Deviations
from this gravitational potential and the presence of differential
rotation, both intimately related to the internal angular momentum
distribution, should be investigated. Indeed, several differential
rotation theories predict surface deformations stronger than that
of uniform rotation by considering that the angular velocity
increases towards the stellar center. Two interesting examples are
"shellular" rotation (Zahn 1992) and laws with angular momentum
constant on cylinders (Bodenheimer 1971). In this context our
result on Achernar's surface distortion should also impact other
internal mechanisms like meridional circulation, turbulence,
transport and diffusion of the chemical elements and angular
momentum, increase of mass loss with rotation as well as
anisotropies in the mass ejection and wind density from rotating
stars (Maeder 1999, Maeder \& Meynet 2000).

Finally, the highly distorted shape of Achernar poses the question
of Be stars rotation rate. As argued by several authors (e.g.
Cassinelli 1987,  Owocki 2003) the formation of out-flowing discs
from Be stars remains their central puzzle, where rapid rotation
is the crucial piece. Struve's (1931) original vision of a
critically rotating star, ejecting material from its equator, has
been discarded in the past by observing that Be stars rotate at
most 70\% or 80\% of their critical velocity (typically $\sim 500$
km/s for a B0V star). However, this statistically observed limit
might be biased by the fact that close to or beyond such
velocities the diagnosis of Doppler-broadened spectral lines fails
to determine the rotation value due to gravity darkening (Owocki
2003, Townsend 1997). We believe that only direct measures of Be
star photospheres by interferometry can overcome the challenge to
prove whether these objects rotate close, to a few percent, of
their critical velocity or not. This have a profound impact on
dynamical models of Be stars CSE formation from rapid rotation
combined to mechanisms like pulsation, radiation pressure of
photospheric hot spots, or expelled plasma by magnetic flares.

\section{Acknowledgments}

A. D. S., P.K. and L.A. acknowledge CAPES-Brazil, ESO and
CNES-France for financial support respectively. The authors
acknowledge support from D.Mourard and his team from OCA-France.
LUAN is supported by UNSA and CNRS. Observations with the VLTI
became possible thanks to the VLTI team. We are grateful to Drs.
N.V.Leister and E.Janot-Pacheco for Achernar's spectra.

{}

\end{document}